\documentclass[fleqn,usenatbib]{mnras}

\usepackage{times}
\usepackage[T1]{fontenc}
\usepackage{graphicx}
\usepackage{amsmath}
\usepackage{amssymb}
\usepackage{flushend}

\newcommand{\Msunnom}{\hbox{$\mathcal{M}^{\rm N}_\odot$}}

\newcommand{\Lsunnom}{\hbox{$\mathcal{L}^{\rm N}_\odot$}}

\newcommand{\Teff}{\ensuremath{T_{\rm eff}}}                      
\newcommand{\logg}{\ensuremath{\log g}}                           
\newcommand{\kms}{\,km\,s$^{-1}$}                                 
\newcommand{\mc}[1]{\multicolumn{2}{c}{#1}}

\newcommand{\gaia}{\textit{Gaia}}

\title[Pulsations in Pleiades eclipsing binary HD\,23642]
      {$\delta$\,Scuti pulsations in the bright Pleiades eclipsing binary HD\,23642}

\author[Southworth et al.]
       {John Southworth\,$^{1}$, S.\ J.\ Murphy$^2$, K.\ Pavlovski$^3$ \\
        $^1$\,Astrophysics Group, Keele University, Staffordshire, ST5 5BG, UK \\
        $^2$\,Centre for Astrophysics, University of Southern Queensland, Toowoomba, QLD 4350, Australia \\
        $^3$\,Department of Physics, Faculty of Science, University of Zagreb, Bijenicka cesta 32, 10000 Zagreb, Croatia
        \vspace*{-15pt}
        }

\date{Accepted XXX. Received YYY; in original form ZZZ}
\pubyear{2022}

\begin{document} \label{firstpage} \pagerange{\pageref{firstpage}--\pageref{lastpage}} \maketitle 

\begin{abstract}
We announce the discovery of pulsations in HD\,23642, the only bright eclipsing system in the Pleiades, based on light curves from the Transiting Exoplanet Survey Satellite (TESS). We measure 46 pulsation frequencies and attribute them to $\delta$\,Scuti pulsations in the secondary component. We find four $\ell=1$ doublets, three of which have frequency splittings consistent with the rotation rate of the star. The dipole mode amplitude ratios are consistent with a high stellar inclination angle and the stellar rotation period agrees with the orbital period. Together, these suggest that the spin axis of the secondary is aligned with the orbital axis. We also determine precise effective temperatures and a spectroscopic light ratio, and use the latter to determine the physical properties of the system alongside the TESS data and published radial velocities. We measure a distance to the system in agreement with the \gaia\ parallax, and an age of $170 \pm 20$\,Myr based on a comparison to theoretical stellar evolutionary models.
\end{abstract}



\section{Introduction}
\label{sec:intro}

The Pleiades is one of the closest and most extensively studied star clusters, containing $\sim$1300 stars at a distance of $136$\,pc \citep{Melis+14sci,Heyl++22apj}. It is a benchmark for studying theoretical modelling \citep[e.g.][]{VandenbergBridges84apj} the kinematics of star clusters \citep{Lodieu+19aa}, 
binarity \citep{Torres++21apj}, pulsations \citep{Murphy+22mn}, rotation \citep{Rebull+16aj}, and the connection between rotation, lithium depletion and stellar inflation \citep{SomersPinsonneault15mn,Bouvier+18aa}.

Eclipsing binary stars (EBs) are another crucial research area for improving our understanding of stellar evolution \citep{Torres++10aarv}. The masses, radii and effective temperatures (\Teff s) of stars can be determined to precisions approaching 0.2\% \citep{Maxted+20mn,Miller++20mn}, allowing their use as checks and calibrators of theoretical models \citep[e.g.][]{ClaretTorres18apj,Tkachenko+20aa}. In particular, EBs in open clusters are enticing targets for multiple scientific goals \citep[e.g.][]{Me++04mn,Brogaard+11aa,Torres+18apj}.

A third phenomenon well suited to extending our understanding of stellar physics is that of pulsations. $\delta$\,Scuti stars pulsate in pressure modes of low radial order ($n \sim 1 \dots 10$) that are mostly sensitive to the stellar envelope \citep{Aerts++10book,Kurtz22araa}. These modes probe the stellar density, making them good indicators of age \citep{Aerts15an}, especially when mass and/or metallicity are already constrained. Recently, the discovery that $\delta$\,Scuti stars pulsate in regular patterns \citep{Bedding+20nat} has allowed pulsation modes to be identified in many stars \citep[e.g.][]{Kerr+22apj,Currie+22}, including five members of the Pleiades \citep{Murphy+22mn}, facilitating detailed asteroseismic modelling. In some cases, this confers an age precision better than 10\% \citep{Murphy+21mn}.


HD\,23642 (V1229\,Tau, HII\,1431) was found to be a double-lined spectroscopic binary by \citet{Pearce57pdao} and \citet{Abt58apj}. Its eclipsing nature was announced independently by \citet{Miles99jbaa} and \citet{Torres03ibvs} using data from the \textit{Hipparcos} satellite. Detailed analyses of the system using ground-based data have been presented by \citet{Griffin95jrasc,Torres03ibvs,Munari+04aa,Me++05aa} and \citet{Groenewegen+07aa}. HD\,23642 was observed in long cadence mode by the K2 satellite \citep{Howell+14pasp} in Campaign 4\footnote{HD\,23642 was requested as a target by nine Guest Observer proposals including GO4028 (PI Southworth) and GO4035 (PI Murphy).}. An analysis of these data was presented by \citet{David+16aj}.

In this work we present the discovery of $\delta$\,Scuti pulsations in this important EB and determine the physical properties of the system to high precision for the first time. We note that independent detections are presented by \citet{Chen+22apjs} without detailed analysis, and by \citet{Bedding+23xxx}. Section\,\ref{sec:obs} in the current work outlines the observations used, Sections \ref{sec:spec} and \ref{sec:lc} present our spectroscopic and photometric analyses, Section\,\ref{sec:puls} is dedicated to the pulsation analysis, and our work is concluded in Section\,\ref{sec:conc}.


\begin{figure*} \includegraphics[width=\textwidth]{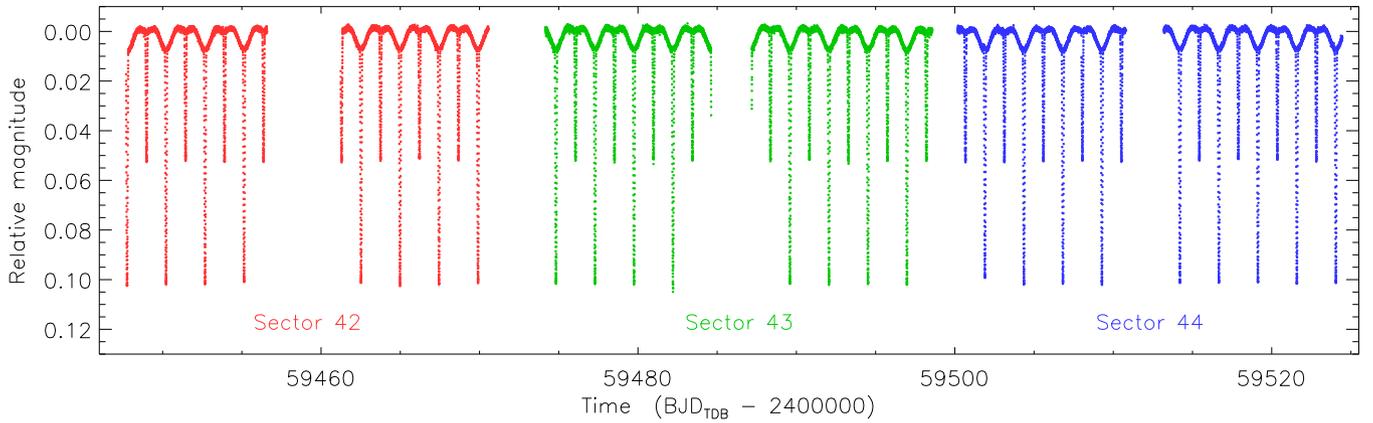}
\caption{\label{fig:tess:time} TESS simple aperture photometry (SAP)
light curve of HD\,23642. The sectors are labelled.} \end{figure*}

\section{Observations}
\label{sec:obs}


HD\,23642 was observed using the TESS mission \citep{Ricker+15jatis_alt} in three consecutive sectors (42--44) covering 76 days (2021/08/20 to 2021/11/06), at a cadence of 120\,s. We downloaded the data from the Mikulski Archive for Space Telescopes (MAST) archive 
and extracted the simple aperture photometry (SAP) from the \textsc{fits} files \citep{Jenkins+16spie}; the PDCSAP data are practically identical. We included only the data with a QUALITY flag of zero, totalling 44\,438 points, and converted them into differential magnitude. The data errors were not used as they were much smaller than the scatter of the measurements. The full TESS data are shown in Fig.\,\ref{fig:tess:time}.

We observed HD\,23642 during two observing runs in November 2006 using the Nordic Optical Telescope (NOT) and its FIbre Echelle Spectrograph (FIES). A total of 27 \'echelle spectra were obtained using the medium-resolution fibre in bundle A
, which covered 364--736\,nm with a resolving power of $R \approx 47\,000$. Exposure times of 600\,s yielded a signal to noise ratio (S/N) of approximately 200, although some spectra had a lower S/N due to cloudy conditions. The data were reduced using {\sc iraf} 
\'echelle package routines, with particular care taken in the normalisation and merging of the \'echelle orders \citep{Kolbas+15mn}.


\section{Spectroscopic analysis}
\label{sec:spec}

We first sought to measure the spectroscopic parameters of the two stars, in particular their \Teff s and light ratio. To do this we applied the method of spectral disentangling \citep{SimonSturm94aa} using the Fourier approach \citep{Hadrava95aas} and concentrating on the 650-670\,nm spectral range which includes the H$\alpha$ line. This wavelength interval is contaminated by telluric lines, which we removed before the disentangling process. We used the {\sc FDbinary} code \citep{Ilijic+04aspc} and our standard methods \citep{PavlovskiHensberge10aspc,Pavlovski++18mn}. We also fixed the velocity amplitudes of the stars to the values measured 
by \citet{Torres++21apj}, so effectively ran in spectral separation mode. 

The H$\alpha$ profiles of the two stars were then modelled to determine the \Teff s and light ratio. We fixed the surface gravities and rotational velocities of the stars to the values determined in Section\,\ref{sec:absdim}, thus avoiding the degeneracy between \Teff\ and $\logg$ present in the Balmer lines of hot stars. This approach required us to iterate our analysis with that described in Section\,\ref{sec:absdim} to ensure internal consistency; the iteration converged within one step. Optimal fitting was performed with the {\sc starfit} code \citep{Kolbas+15mn}, which uses a genetic algorithm to search for the best fit within a grid of synthetic spectra pre-calculated using the {\sc uclsyn} code \citep{Smalley++01}. The fractional light contributions of the two components were forced to sum to unity \citep{Tamajo++11aa} and only the wings of the H$\alpha$ line were fitted (with metallic lines masked). The uncertainties were calculated using the MCMC approach described in \citet{Pavlovski++18mn}.

We found \Teff s of $10\,200 \pm 90$~K and $7670 \pm 85$~K for the two stars, and a light ratio of $\ell_{\rm B}/\ell_{\rm A}{\rm(H}\alpha) = 0.313 \pm 0.010$. HD\,23642\,A is known to be chemically peculiar: \citet{AbtLevato78pasp} classified its spectrum as A0Vp(Si) + Am. Our light ratio should not be affected by this because it was obtained from only the H$\alpha$ line. We propagated it to the TESS passband using BT-Settl theoretical spectra \citep{Allard++12rspta} to obtain $\ell_{\rm B}/\ell_{\rm A}{\rm(TESS)} = 0.328 \pm 0.011$ where the errorbar includes the contributions from $\ell_{\rm B}/\ell_{\rm A}{\rm(H}\alpha)$ and both \Teff s.


\begin{figure*} \includegraphics[width=\textwidth]{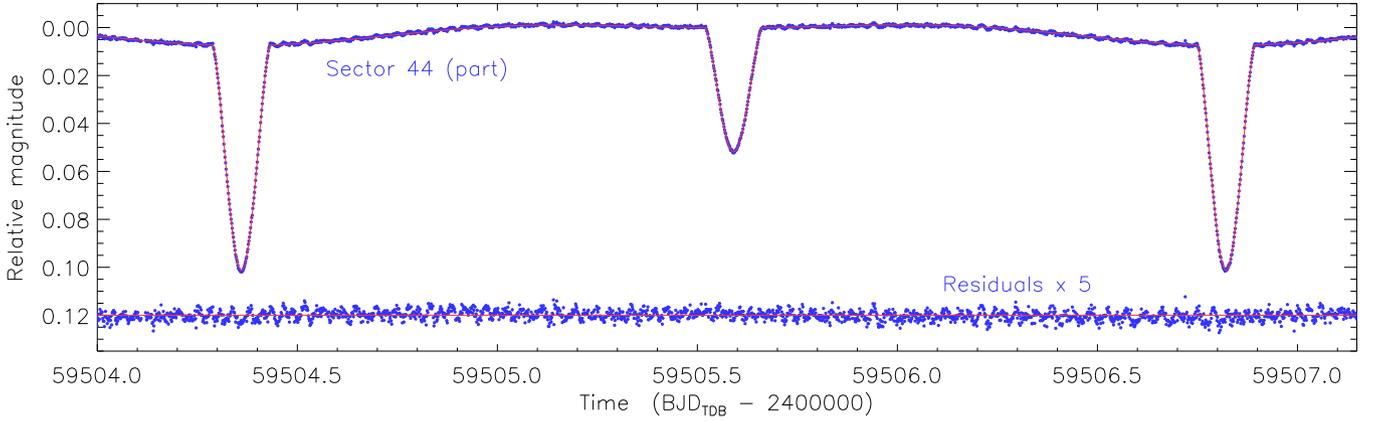}
\caption{\label{fig:tess:bit} A short section of the TESS light curve, chosen at random, is shown (blue points)
along with the {\sc jktebop} best fit (red line). The residuals are displayed offset to the
base of the figure and magnified by a factor of 5 to make the pulsations visible.} \end{figure*}

\section{Light curve and physical properties}
\label{sec:lc}
\label{sec:absdim}

The light curve of HD\,23642 shows shallow partial eclipses and clear reflection and ellipsoidal effects. The stars are well-separated and the system is suitable for analysis with the {\sc jktebop} 
code \citep{Me++04mn2,Me13aa}, for which we used version 43. We fitted for the fractional radii of the stars in the form of their sum ($r_{\rm A}+r_{\rm B}$) and ratio ($k = {r_{\rm B}}/{r_{\rm A}}$), the orbital inclination ($i$), the central surface brightness ratio of the two stars ($J$), the amount of third light ($L_3$), the reference time of mid-eclipse ($T_0$), and the orbital period ($P$). Limb darkening was included using the power-2 law \citep{Hestroffer97aa} with the scaling coefficient for each star ($c_1$ and $c_2$) fitted and the power-law coefficients ($\alpha_1$ and $\alpha_2$) fixed at values from \citet{ClaretSouthworth22aa}. We also included one quadratic function per TESS half-sector to normalise the light curve to zero differential magnitude, to allow for the possibility of slow drifts in brightness for either instrumental or astrophysical reasons. The pulsations are of much lower amplitude than the eclipses so were treated as red noise. A circular orbit was assumed.

Our initial solutions gave measurements of the fractional radii to a disappointing precision. This is caused by the ratio of the radii being poorly determined for shallow partial eclipses, a well-known phenomenon that has been noted before for this system \citep{Me++05aa,David+16aj}. We therefore imposed the spectroscopic light ratio from Section\,\ref{sec:spec} as a Gaussian prior in our solution. This significantly improved the precision of the fitted parameters, and is the only viable approach in a system like this with shallow partial eclipses and a noise limit set by the presence of pulsations. Uncertainties in the fitted parameters were determined by Monte Carlo and residual-permutation simulations \citep{Me08mn}, taking the larger of the two options for each quantity. The values and uncertainties of the parameters are given in Table\,\ref{tab:absdim}. We have added an extra uncertainty of $\pm$0.0010 in quadrature to the uncertainties in $r_{\rm A}$ and $r_{\rm B}$ to account for the variations in these parameters between different model choices, specifically about whether or not to fit for $L_3$ or limb darkening.

\begin{table} \centering
\caption{\label{tab:absdim} Properties of the HD\,23642 system. The velocity amplitudes $K$ are from \citet{Torres++21apj}.}
\begin{tabular}{lr@{\,$\pm$\,}lr@{\,$\pm$\,}l}
\hline
Quantity and unit                   & \mc{Star~A} & \mc{Star~B} \\
\hline
\multicolumn{5}{l}{\it Spectroscopic parameters:} \\
$K$ (\kms)                          & 100.07 & 0.23 & 142.60 & 0.28 \\
\Teff\ (K)                          & 10200 & 90 & 7670 & 85 \\
Light ratio at H$\alpha$            & \multicolumn{4}{c}{$0.313 \pm 0.010$} \\
Light ratio in TESS passband        & \multicolumn{4}{c}{$0.328 \pm 0.011$} \\
\multicolumn{5}{l}{\it {\sc jktebop} analysis:} \\
$r_{\rm A}+r_{\rm B}$               & \multicolumn{4}{c}{$0.2664 \pm 0.0011$} \\
$k$                                 & \multicolumn{4}{c}{$0.784 \pm 0.012$} \\
$i$ ($^\circ$)                      & \multicolumn{4}{c}{$78.63 \pm 0.09$} \\
$J$                                 & \multicolumn{4}{c}{$0.5561 \pm 0.0011$} \\
$\ell_3$                            & \multicolumn{4}{c}{$0.029 \pm 0.014$} \\
$c$                                 & \mc{0.52 fixed}  & \mc{0.63 fixed} \\
$\alpha$                            & \mc{0.45 fixed}  & \mc{0.42 fixed} \\
$P$ (d)                             & \multicolumn{4}{c}{$2.46113223 \pm 0.00000060$} \\
$T_0$ (BJD$_{\rm TDB}$)             & \multicolumn{4}{c}{$2459509.282357 \pm 0.000007$} \\
Fractional radii                    & 0.1494 & 0.0016 & 0.1171 & 0.0017 \\
\multicolumn{5}{l}{\it Physical properties:} \\
Mass (\Msunnom)                     & 2.273 & 0.011 & 1.595 & 0.008 \\
Radius (\Msunnom)                   & 1.799 & 0.019 & 1.410 & 0.021 \\
\logg\ (c.g.s.)                     & 4.285 & 0.009 & 4.342 & 0.13 \\
$V_{\rm synch}$ (\kms)              & 36.98 & 0.40  & 28.99 & 0.42  \\
Luminosity $\log(L/\Lsunnom)$       & 1.499 & 0.018 & 0.792 & 0.023 \\
Reddening $E_{B-V}$ (mag)           & \multicolumn{4}{c}{$0.040 \pm 0.010$} \\
$K$-band distance (pc)              & \multicolumn{4}{c}{$134.7 \pm 2.0$} \\
\hline
\end{tabular}
\end{table}



We determined the physical properties of the system from the results of the {\sc jktebop} analysis and the velocity amplitudes of the stars measured by \citet{Torres++21apj}. This was done using the {\sc jktabsdim} code \citep{Me++05aa} modified to report results on the IAU scale \citep{Prsa+16aj}. The full set of measured system properties are given in Table\,\ref{tab:absdim}. Our masses are in good agreement with those found by other authors, with minor differences due to the newer velocity amplitudes adopted. The rotational velocities of the stars determined by \citet{Me++05aa}, $37 \pm 2$ and $33 \pm 3$\kms, are in agreement with the synchronous values, suggesting that both stars are rotating synchronously.

The measured radii, however, are significantly different from previous values \citep{Munari+04aa,Me++05aa,Groenewegen+07aa,David+16aj} in the sense that $R_{\rm A}$ is larger and $R_{\rm B}$ is smaller. This solves an existing problem with HD\,23642\,B, which was previously found to be significantly larger than predicted by theoretical evolutionary models for the Pleiades' age and metallicity. We attribute this change to differences in the spectroscopic light ratios adopted in those studies, which are slightly larger than the one measured in the current work. Both components of HD\,23642 are known to be chemically peculiar so light ratios determined from metal lines are unreliable. Our new light ratio should be preferred because it is based on the H$\alpha$ line so is not affected by the chemical peculiarity, and is also close to the wavelengths transmitted by the TESS passband. A visual comparison of the masses, radii and \Teff s of the primary star to predictions from the {\sc parsec} evolutionary models \citep{Bressan+12mn} for solar metallicity shows good agreement for an age of $170 \pm 20$\,Myr. We note that this is at the upper limit of accepted ages of the Pleiades cluster \citep{Gossage+18apj,Murphy+22mn}. The secondary component is 1.5$\sigma$ smaller than expected: further investigation is needed to understand this.

To determine the distance ($d$) and interstellar reddening ($E_{B-V}$) to the system we used the $BV$ apparent magnitudes from the Tycho satellite \citep{Hog+00aa}, the $JHK_s$ apparent magnitudes from 2MASS \citep{Skrutskie+06aj} converted into the Johnson system using transformations from \citet{Carpenter01aj}, and bolometric corrections from \citet{Girardi+02aa}. We determined the value of $E_{B-V}$ that results in consistent distances across the $BVJHK$ bands, finding $E_{B-V} = 0.040 \pm 0.010$ and $d = 134.7 \pm 2.0$~pc. This distance is slightly shorter than the $138.3 \pm 0.1$~pc determined by simple inversion of the parallax from \textit{Gaia} EDR3 \citep{Gaia21aa}. This $E_{B-V}$ is specific to HD\,23642, is consistent with previous studies \citep{Taylor08aj}, and is not affected by reddening variations between Pleiades members.


\begin{figure*}
\begin{center}
\includegraphics[width=\textwidth]{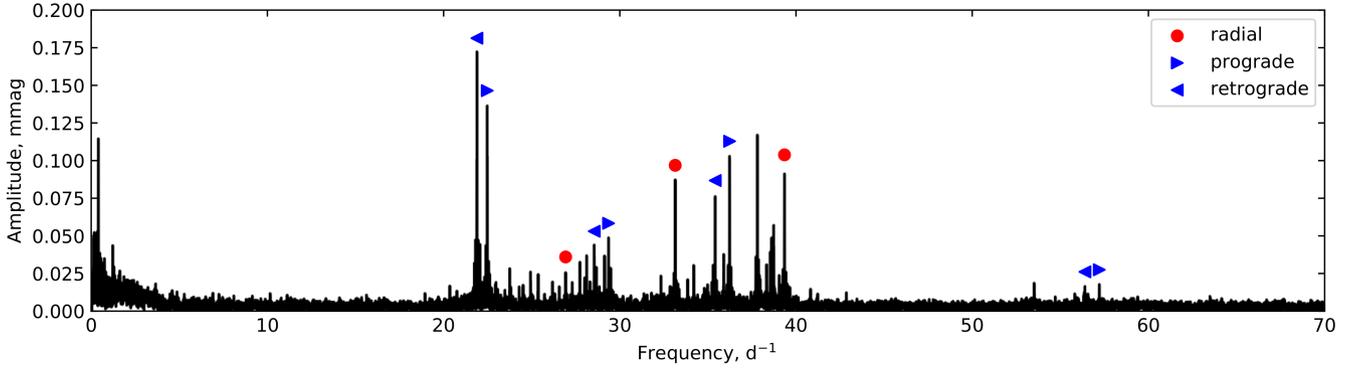}
\caption{\label{fig:FT} Fourier transform of the light curve residuals, after subtracting the eclipse
model, revealing several pulsations at the 0.1\,mmag level. Mode identifications (Section\,\ref{sec:puls})
are overlaid, showing radial modes as well as a series of rotationally split dipole modes.}
\end{center}
\end{figure*}

\section{Analysis of the pulsations}
\label{sec:puls}

The residual light curve after subtraction of the {\sc jktebop} model shows several pulsation modes at the 0.1\,mmag level (Fig.\,\ref{fig:FT}). We used the {\sc period04} code \citep{LenzBreger04iaus} to extract 46 pulsations at frequencies $f>20$~d$^{-1}$ down to 0.015\,mmag amplitude, and used non-linear least squares to optimise the frequencies. The table of frequencies is given in the Appendix. From the frequencies and \Teff s of the stars we deduce that they represent $\delta$\,Scuti pulsations in the secondary component, as the primary is hotter than the $\delta$\,Scuti instability strip.

\begin{figure}
\begin{center}
\includegraphics[width=0.48\textwidth]{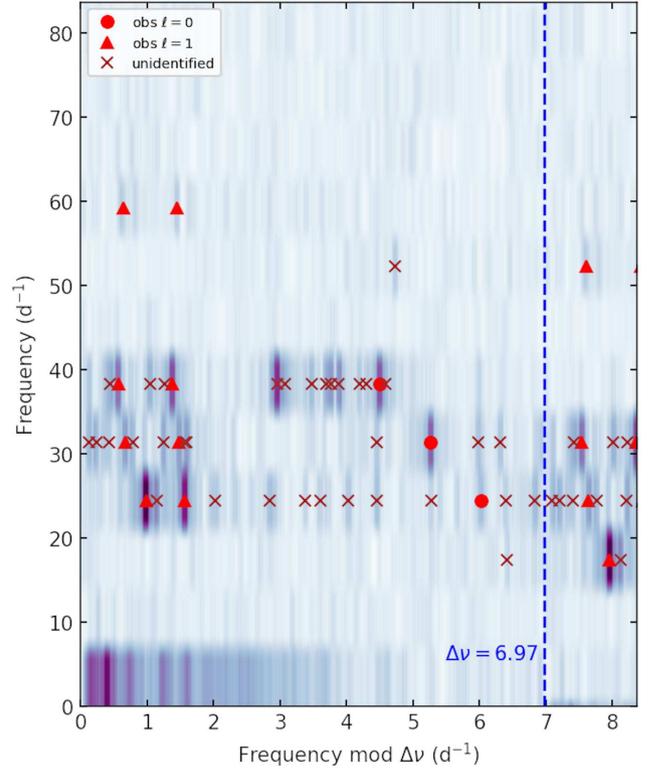}
\caption{\label{fig:echelle} \'Echelle diagram for HD\,23642, marked with mode identifications. 
Radial modes are shown as circles, dipole modes are triangles. Modes without an identifications are shown as
crosses, some of which may belong to the other star; most will be modes of higher degree ($\ell\geq2$).}
\end{center}
\end{figure}

We used the {\sc echelle} package \citep{HeyBall20zenodo} to manually find values of the asteroseismic large spacing, $\Delta\nu$, that give vertical patterns in an \'echelle diagram. At $\Delta\nu=6.97$\,d$^{-1}$, there are two parallel ridges on the left-hand side of the \'echelle (Fig.\,\ref{fig:echelle}), at an $x$-location suggestive of $\ell=1$ modes \citep{Bedding+20nat,Murphy+21mn}. With the same $\Delta\nu$, part of a radial mode series can also be identified, with period ratios consistent with low-order radial modes \citep{Netzel+22mn}. It is noteworthy that the independently-determined $\Delta\nu$ is consistent with the value of five other $\delta$\,Scuti stars in the Pleiades \citep[6.82--6.99\,d$^{-1}$;][]{Murphy+22mn}, and also suggests that the system is relatively young ($\lesssim$200\,Myr).

Since $\ell=1$ doublets are seen, which are interpretable as $m=\pm1$ pairs, the inclination of the pulsating star is not small \citep{GizonSolanki03apj}. However, at some orders, most notably at $n=3$, there is a peak slightly offset from halfway between the two identified $\ell=1$ modes that could be a central component, after second-order effects of rotation are accounted for. To evaluate this possibility would require the calculation of rotating evolutionary and pulsation models. If confirmed, it implies that the stellar inclination is also not completely edge-on and, given the measured orbital inclination, implies that the spin and orbital axes of the pulsating star are aligned.

The identification of four $\ell=1$ doublets offers the chance of a consistency check on the mode identification. If the doublets all have approximately the same splitting, it strengthens the proposed mode identifications. For the doublets at $n=6$, 3, 2, and 1 we measure splittings of 0.813, 0.813, 0.812, and 0.586\,d$^{-1}$, respectively. Thus, the $n=6$, 3, and 2 doublets seem secure, but the $n=1$ doublet does not. Since the Ledoux constant, $C_{\rm n,\ell}$, is close to zero for p modes, we can estimate the stellar rotation rate, $\Omega$, to first order based on these frequency splittings \citep{Aerts++10book}:
\begin{eqnarray}
\nu_{n,\ell,m} = \nu_0 + m\Omega(1 - C_{n,\ell}),
\end{eqnarray}
where $\nu_0$ is the rest-frame frequency of the pulsation mode (we adopt the convention that prograde modes have positive $m$). The stellar rotation frequency is therefore 0.41\,d$^{-1}$, corresponding to a rotation period of 2.46\,d. This is equal to the orbital period of the system, and the stars are known to rotate approximately synchronously (see Section\,\ref{sec:absdim}), so this supports our inference of the stellar rotational inclination.


\section{Summary and conclusions}
\label{sec:conc}

HD\,23642 has it all: youth, eclipses, pulsations, chemical peculiarity, and membership of the Pleiades. We establish its physical properties to high precision for the first time, helped particularly by the measurement of a spectroscopic light ratio immune to the chemical peculiarity. We determine a distance and interstellar reddening to the system in good agreement with previous measurements, and infer an age of $170 \pm 20$\,Myr for the Pleiades by comparison to {\sc parsec} evolutionary model predictions.

A total of 46 pulsation frequencies are detected to high significance from the three consecutive sectors of TESS photometry. We attribute them to $\delta$\,Scuti pulsations in the secondary star. We use an \'echelle diagram to assign modes to 11 of the pulsations, based on the identification of $\ell=1$ doublets. An asteroseismic large spacing of $\Delta\nu=6.97$\,d$^{-1}$ allows identification of a series of radial modes with period ratios consistent with other $\delta$\,Scuti stars in the Pleiades. The stellar rotation rate we find from the mode splittings is in agreement with the spectroscopic $v\sin i$ and the orbital period of the system. This implies that the spin axis of the pulsating secondary component is aligned with the orbital axis of the system. HD\,23642 is well suited to detailed analysis with evolutionary and pulsation models including rotation.


\vspace*{-7pt}
\section*{Data availability}

The TESS data used in this work are available in the MAST archive (https://mast.stsci.edu/portal/Mashup/Clients/Mast/Portal.html). The FIES spectra are available in reduced form from the NOT archive (http://www.not.iac.es/archive/).

\vspace*{-7pt}
\section*{Acknowledgements}

We thank Dominic Bowman and Hiromoto Shibahashi for useful discussions.
The TESS data presented in this paper were obtained from the Mikulski Archive for Space Telescopes (MAST) at the Space Telescope Science Institute (STScI). 
STScI is operated by the Association of Universities for Research in Astronomy, Inc. 
Support to MAST for these data is provided by the NASA Office of Space Science. 
Funding for the TESS mission is provided by the NASA Explorer Program.
SJM was supported by the Australian Research Council through Future Fellowship FT210100485.


\vspace*{-7pt}
\bibliographystyle{mnras}


\appendix         

\section{Measured pulsation frequencies in HD\,23642}

\begin{table}
\centering
\caption{\label{tab:freqs} Extracted frequencies with corresponding amplitudes
and provisional mode identifications for the pulsating component of HD\,23642.}
\begin{tabular}{ccccr}
\hline
$f$ (d$^{-1}$) & amplitude (mmag) & $n$ & $\ell$ & $m$ \\
\hline
57.2112 & 0.0175 &  6 &  1 &   1  \\
56.3981 & 0.0161 &  6 &  1 & $-1$ \\
36.2304 & 0.1029 &  3 &  1 &   1  \\
35.4177 & 0.0768 &  3 &  1 & $-1$ \\
29.3618 & 0.0483 &  2 &  1 &   1  \\
28.5496 & 0.0431 &  2 &  1 & $-1$ \\
22.4772 & 0.1365 &  1 &  1 &   1  \\
21.8913 & 0.1713 &  1 &  1 & $-1$ \\
39.3467 & 0.0938 &  4 &  0 &   0  \\
33.1444 & 0.0869 &  3 &  0 &   0  \\
26.9261 & 0.0260 &  2 &  0 &   0  \\
53.5132 & 0.0188 & -- & -- &  -- \\
39.4291 & 0.0302 & -- & -- &  -- \\
39.1434 & 0.0157 & -- & -- &  -- \\
39.0450 & 0.0199 & -- & -- &  -- \\
38.7352 & 0.0514 & -- & -- &  -- \\
38.6162 & 0.0407 & -- & -- &  -- \\
38.5353 & 0.0362 & -- & -- &  -- \\
38.3290 & 0.0296 & -- & -- &  -- \\
37.9217 & 0.0272 & -- & -- &  -- \\
37.8036 & 0.1171 & -- & -- &  -- \\
36.1084 & 0.0165 & -- & -- &  -- \\
35.8956 & 0.0370 & -- & -- &  -- \\
35.2961 & 0.0289 & -- & -- &  -- \\
34.1939 & 0.0286 & -- & -- &  -- \\
33.8544 & 0.0206 & -- & -- &  -- \\
32.3307 & 0.0221 & -- & -- &  -- \\
29.4768 & 0.0239 & -- & -- &  -- \\
29.4484 & 0.0198 & -- & -- &  -- \\
29.1269 & 0.0368 & -- & -- &  -- \\
28.6681 & 0.0302 & -- & -- &  -- \\
28.3104 & 0.0203 & -- & -- &  -- \\
28.1144 & 0.0378 & -- & -- &  -- \\
28.0038 & 0.0230 & -- & -- &  -- \\
27.7391 & 0.0347 & -- & -- &  -- \\
27.3022 & 0.0181 & -- & -- &  -- \\
26.1835 & 0.0203 & -- & -- &  -- \\
25.3700 & 0.0252 & -- & -- &  -- \\
24.9365 & 0.0264 & -- & -- &  -- \\
24.5100 & 0.0159 & -- & -- &  -- \\
24.2900 & 0.0160 & -- & -- &  -- \\
23.7500 & 0.0281 & -- & -- &  -- \\
22.9355 & 0.0181 & -- & -- &  -- \\
22.0604 & 0.0403 & -- & -- &  -- \\
20.3501 & 0.0157 & -- & -- &  -- \\
\hline
\end{tabular}
\end{table}

\bsp \label{lastpage} \end{document}